\documentclass[aps,prl,reprint,superscriptaddress]{revtex4-2}

\usepackage{graphicx}
\usepackage{epstopdf}
\usepackage{amssymb}
\usepackage{xcolor}

\begin{document}

\def\cm{cm$^{-1}$}
\def\CuCN{$\kappa$-(BEDT\--TTF)$_2$\-Cu$_2$(CN)$_3$}
\def\AgCN{$\kappa$-(BEDT\--TTF)$_2$\-Ag$_2$(CN)$_3$}
\def\STF{$\kappa$-[(BEDT\--TTF)$_{1-x}$\-(BEDT\--STF)$_{x}$]$_2$\-Cu$_2$(CN)$_3$}
\def\stf{$\kappa$\--STF$_x$}
\def\dmit{$\beta^{\prime}$\--EtMe$_3$\-Sb[Pd(dmit)$_2$]$_2$}
\def\BrCl{$\kappa$-(BEDT\--TTF)$_2$\-Cu[N(CN)$_2$]Br$_x$Cl$_{1-x}$}
\def\Cl{$\kappa$-(BEDT\--TTF)$_2$\-Cu[N(CN)$_2$]Cl}
\def\Br{$\kappa$-(BEDT\--TTF)$_2$\-Cu[N(CN)$_2$]Br}
\def\HgCl{$\kappa$-(BEDT\--TTF)$_2$\-Hg(SCN)$_2$Cl}
\def\Mn{$\kappa$-(BETS)$_2$\-Mn[N(CN)$_2$]$_3$}

\title{Mott Intermittency at the Metal–Insulator Boundary}

\author{Yuxin Wang}
\affiliation{National High Magnetic Field Laboratory, Florida State University, Tallahassee, Florida 32310, USA}
\affiliation{Department of Physics, Florida State University, Tallahassee, Florida 32306, USA}

\author{Vladimir Dobrosavljevi\'{c}}
\affiliation{National High Magnetic Field Laboratory, Florida State University, Tallahassee, Florida 32310, USA}
\affiliation{Department of Physics, Florida State University, Tallahassee, Florida 32306, USA}

\author{Jan Jaroszy\'nski}
\affiliation{National High Magnetic Field Laboratory, Florida State University, Tallahassee, Florida 32310, USA}

\author{Yohei Saito}
\affiliation{Department of Physics, Hokkaido University, Sapporo, Hokkaido 060-0810, Japan}

\author{Atsushi Kawamoto}
\affiliation{Department of Physics, Hokkaido University, Sapporo, Hokkaido 060-0810, Japan}

\author{Andrej Pustogow}
\affiliation{Institute of Solid State Physics, Technische Universit\"{a}t Wien, Vienna A-1040, Austria}

\author{Martin Dressel}
\affiliation{1.~Physikalisches Institut, Universit\"{a}t Stuttgart, 70569 Stuttgart, Germany}

\author{Dragana Popovi\'{c}}
\email[]{dragana@magnet.fsu.edu}
\affiliation{National High Magnetic Field Laboratory, Florida State University, Tallahassee, Florida 32310, USA}
\affiliation{Department of Physics, Florida State University, Tallahassee, Florida 32306, USA}

\date{February 24, 2026}

\begin{abstract}

The resistivity maximum at a temperature $T=T_{\mathrm{max}}$ is a recurring feature of bandwidth-tuned Mott systems, yet its meaning remains controversial: is it a coherence-incoherence crossover of an electronically homogeneous metal, or does it mark the onset of transport through a mixed landscape of metallic and insulating regions? Even more debated is whether a true phase-coexistence regime survives in the relevant parameter range, or whether apparent inhomogeneity is merely extrinsic. Here we address these questions by moving beyond temperature sweeps and probe charge transport in the time domain. Near $T=T_{\mathrm{max}}$, we find that the resistance of a model system, a quasi-two-dimensional Mott spin liquid material, exhibits clear random-telegraph switching between discrete levels over long timescales. The statistics of the switching -- sharp two-level behavior with thermally activated dwell times -- point to a mesoscopic “current-controlling” region that dynamically toggles between metallic and insulating states, intermittently opening and closing the dominant conduction channel. This characteristic fluctuating dynamics provides direct evidence for intrinsic metal–insulator coexistence and establishes $T \sim T_{\mathrm{max}}$ as the regime of Mott intermittency, where transport is governed by stochastic domain switching rather than quasiparticle decoherence.

\end{abstract}

\maketitle

The interaction-driven metal-insulator transition (MIT) \cite{Mott_MIT1990}  marks the boundary of the conventional metallic regime, where the fundamental nature of elementary excitations changes dramatically. While being recognized as one of the key open questions in the physics of electronic systems, the physical nature of the transition region has long remained the subject of controversy and debate \cite{Tan_Dobrosavljevic_Crystals2022}. New insight has started to emerge in recent years, with the rise of several experimental platforms that permit systematic and precise control of the approach to the transition point, as well as significant advances in the relevant characterization methods. This included studies of various high-mobility two-dimensional electron gases (2DEGs) in conventional semiconductors \cite{Shashkin_Melnikov_PRB2020}, organic spin-liquid materials \cite{Pustogow_Bories_NatMater2018}, and more recently several transition-metal dichalcogenide moir\'e bilayers \cite{Li_Jiang_Nature2021}. Remarkably, the qualitative transport signatures found in all these systems proved similar, despite the fact that their characteristic (Fermi) energies differ by orders of magnitude, hinting at a surprising degree of universality across all these platforms, with a robust underlying mechanism for the transition. 

In contrast, the spectrum of proposed theoretical scenarios \cite{Tan_Dobrosavljevic_Crystals2022} remains surprisingly broad, which calls for more precise validation against experiments. In this regard, one prominent feature of transport stands out, which is the significant drop of the resistivity ($\rho$) on the metallic side, below the resistivity maxima at temperature $T = T_{\mathrm{max}}$ \cite{Radonjic_Tanaskovic_PRB2012, Moon_Han_PRB2020, Shashkin_Melnikov_PRB2020}. Remarkably, $T_{\mathrm{max}}$ is found to universally decrease towards the transition in all materials, although its origin is attributed to very different physical processes within the theoretically proposed scenarios. For example, in the context of 2DEGs, it was attributed to a percolative mechanism~\cite{spivakRMP} in analogy to colossal-magnetoresistance manganites~\cite{manganitesRMP2001}, to competing ``quantum corrections'' within the weakly disordered Fermi liquid picture~\cite{finkelstein}, or alternatively to the onset of quasiparticle coherence within the Wigner-Mott scenario \cite{wigner-mott,Radonjic_Tanaskovic_PRB2012}, in analogy to heavy fermion behavior \cite{heavyfermion1984RMP}. Of course, these physical pictures dramatically differ in the assumed role of disorder and spatial inhomogeneity, an issue that is difficult to directly access using bulk probes. 

In this regard, quasi-2D organic spin-liquid materials  \cite{Pustogow_Bories_NatMater2018} seem to provide new clues. In these systems, one finds both a uniform correlated phase further on the metallic side and, as demonstrated by dielectric measurements \cite{Pustogow_Rosslhuber_npjqm2021}, a more spatially inhomogeneous phase coexistence region around the transition point. Remarkably, very similar resistivity maxima are found in both regimes, with $T_{\mathrm{max}}$ decreasing towards the transition. This opens the possibility to conclusively pinpoint the relevant transport mechanism in each case. Time-resolved transport, i.e., resistance noise measurements 
 \begin{figure*}
 \includegraphics[width=2\columnwidth]{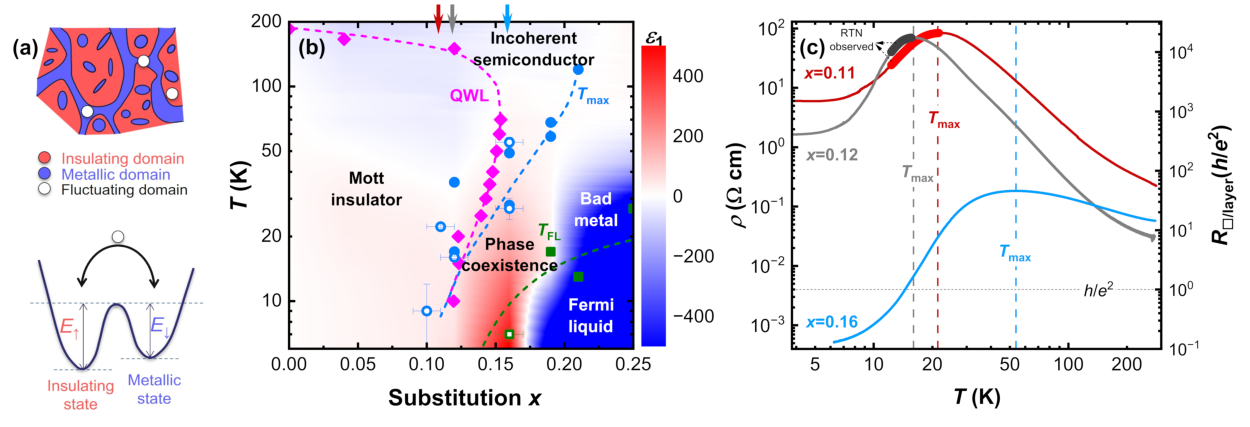}
\caption{(a) Sketch of domain fluctuations in a metal-insulator mixture at the percolation threshold.
(b) $T-x$ phase diagram of \stf~where the color scale maps the dielectric permittivity $\epsilon_1$~\cite{Pustogow_Rosslhuber_npjqm2021}. Magenta diamonds, blue circles, and green squares represent the quantum Widom line (QWL), $T_\mathrm{max}$, and the Fermi liquid temperature ($T_\mathrm{FL}$), respectively; adapted from refs.~\cite{Pustogow_Rosslhuber_npjqm2021, Pustogow_Saito_NatComm2021,Pustogow_Bories_NatMater2018} (solid symbols) and~\cite{Wang_Dobrosavljevic_AXV2025} (open symbols). The three arrows mark the three $x$ values investigated in this work.
(c) $T$-dependence of $\rho$ (left $y$-axis) and $R_{\square/\mathrm{layer}}$ (right $y$-axis) of \stf~at $x=0.11$ (red), $0.12$ (gray), and $0.16$ (blue), with a peak at $T=T_\mathrm{max}$. The highlighted curves near $T_\mathrm{max}$ indicate the $T$ ranges over which RTN has been observed from the $x=0.11$ and $0.12$ samples.
\label{Fig:intro}}
 \end{figure*}
provide a uniquely sensitive probe for this task.
First, slow domain dynamics are expected in the phase coexistence region of a first-order transition. Coexisting phases have nearly equal chemical potentials, allowing a domain to thermally fluctuate between the two phases, provided that $T$ is sufficiently high to overcome the energy barrier. If the two phases possess distinct transport properties, as in the Mott MIT, domain fluctuations should manifest themselves as two-level switching (TLS) of resistance. Second, such a TLS event is sensitive to the percolation threshold. Away from the threshold, a global transport measurement senses fluctuations from numerous domains throughout the sample, the superposition of which tends to wash out the discrete states from TLS, and typically results in noise with a Gaussian amplitude distribution due to the central limit theorem, provided that the fluctuating domains are uncorrelated. On the other hand, at the percolation threshold, [Fig.~\ref{Fig:intro}(a)], an infinite percolating cluster spanning the entire sample carries almost all charge transport. As a result, very few fluctuating domains that fall into the percolating cluster can generate well-defined discrete states of the global resistance~\cite{Raquet_Anane_PRL2000}. In addition, resistance noise can be measured using the same setup as charge transport, allowing the noise and the maxima in $\rho$ 
to be determined precisely from the same samples, thereby yielding results robust to uncertainties in $T_\mathrm{max}$ arising from sample-to-sample variations.

We study electrical transport and noise properties of \STF~(abbreviated~\stf), a chemically substituted series of quasi-2D half-filled organic charge-transfer salts, where the BEDT\--STF substitution extends the donor-molecular orbital and increases the bandwidth that tunes the Mott transition [Fig.~\ref{Fig:intro}(b)]~\cite{Saito_Lohle_Crystals2021, Saito_Rosslhuber_JMCC2021, supp}. This Mott transition is considered genuine because there are no charge or magnetic orders accompanying the transition~\cite{Dressel_Tomic_AP2020}, which often complicate noise studies (e.g., Refs.~\cite {Thomas_Agarmani_npjs2024, Muller_Thomas_Crystals2018}). In this system, the phase-coexistence region below the Mott critical concentration $x\approx 0.12$ has been well characterized by a strongly enhanced dielectric permittivity [Fig.~\ref{Fig:intro}(b)]~\cite{Pustogow_Rosslhuber_npjqm2021}. 
Our samples are within the phase-coexistence region: $x=0.11$, $0.12$, and $0.16$, with an estimated uncertainty of $\Delta x\sim0.01$~\cite{Wang_Dobrosavljevic_AXV2025}.
We find well-defined random telegraph noise (RTN) near $T_\mathrm{max}$ in both $x=0.11$ and $0.12$, 
but not in the more metallic sample ($x=0.16$). We demonstrate that this RTN aligns with slow domain dynamics, strongly 
supporting the percolation picture in which
$T_\mathrm{max}$ is essentially determined by the percolation threshold~\cite{Tan_Dobrosavljevic_Crystals2022}. 

Measurements were performed along the $c$-axis (the most conductive direction within the 2D plane~\cite{Pinteric_Culo_PRB2014}) of millimeter-sized plate-like single crystals of \stf. Effects of heating, aliasing, and spurious noise were carefully examined and eliminated~\cite{supp}. 
We find that 
$\rho(T)$ exhibits 
a maximum 
at $T=T_\mathrm{max}$ [Fig.~\ref{Fig:intro}(c)], which is 
more pronounced for more resistive samples at lower $x$, consistent with the literature~\cite{Saito_Lohle_Crystals2021}. 
$T_\mathrm{max}\approx53.8~\mathrm{K}$ for $x=0.16$ is markedly higher than for 
$x=0.11$ ($T_\mathrm{max}\approx22.3~\mathrm{K}$) and 
$x=0.12$ ($T_\mathrm{max}\approx16~\mathrm{K}$) samples. The slight nonmonotonic variation of $T_\mathrm{max}$ between the $x=0.11$ and $x=0.12$ samples is 
attributed to the uncertainty in the nominal vlue of $x$.
Notably, while all samples exhibit metallic behavior ($d\rho/dT>0$) at $T<T_\mathrm{max}$, only $\rho(T)$ 
for $x=0.16$ falls below the Mott–Ioffe–Regel (MIR) limit, with the sheet resistance per layer $R_{\square/\mathrm{layer}}\equiv\rho/d<h/e^2$ 
($d\approx 16$~\AA\ is the inter-layer distance~\cite{Saito_Rosslhuber_JMCC2021}, $h$ is the Planck constant, and $e$ is the elementary charge). For 
$x=0.11$ and $x=0.12$, $\rho(T)$
saturates at values that are about 
two orders of magnitude higher than the MIR limit, before the onset of superconducting fluctuations 
at $2$--$3~\mathrm{K}$~\cite{Wang_Dobrosavljevic_AXV2025, Saito_Lohle_Crystals2021}. 
Such a high residual $\rho$ is 
largely attributed to the small volume of the percolating cluster that carries most charge transport \cite{Wang_Dobrosavljevic_AXV2025}.

\begin{figure}[t]
 \includegraphics[width=1\columnwidth]{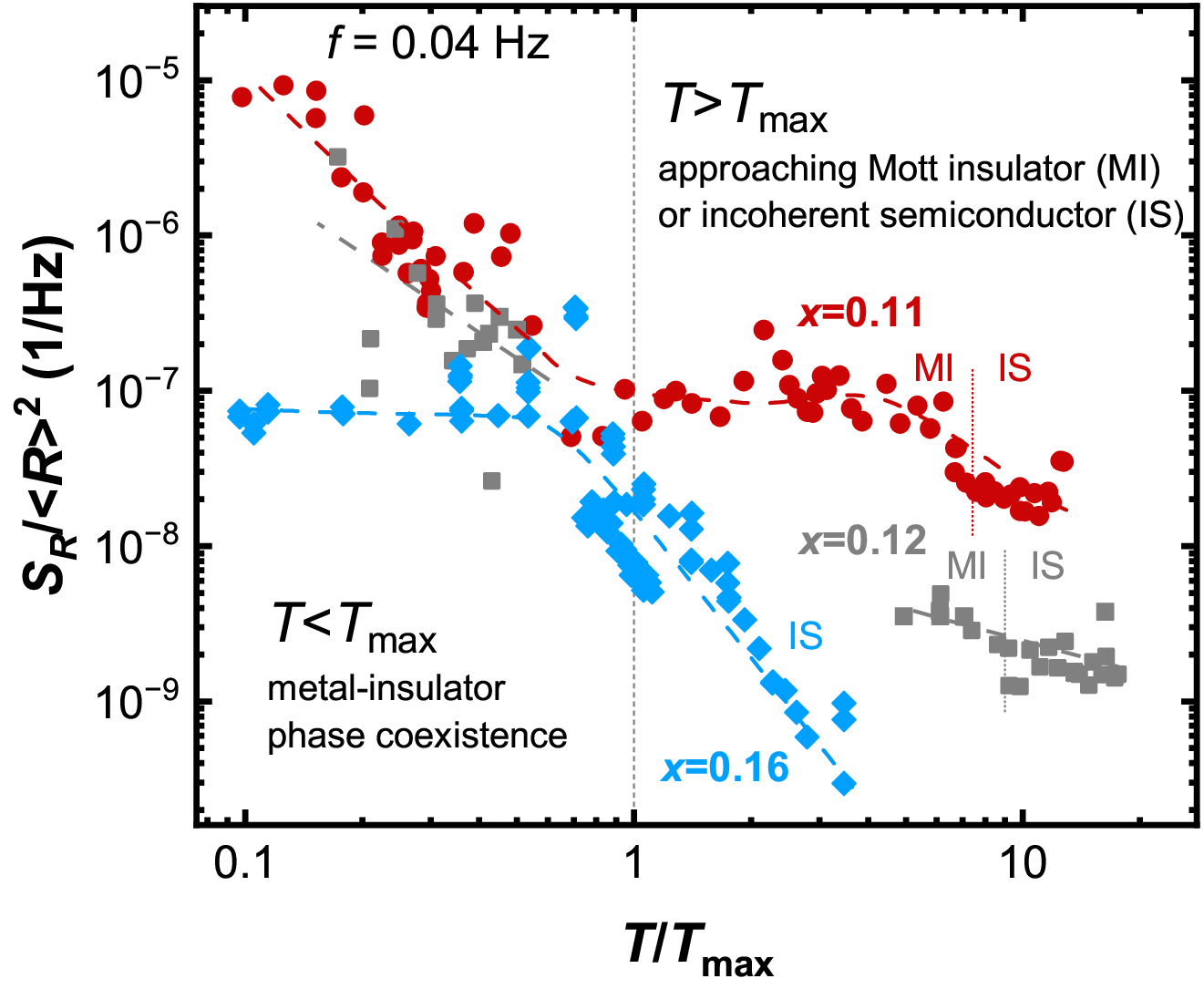}
\caption{The normalized 
power spectra 
$S_R/\langle R \rangle^2$ of the $1/f$ noise (see Supplemental Material, Fig.~S1~\cite{supp})
at $f=0.04~\mathrm{Hz}$ vs 
$T/T_\mathrm{max}$ for $x=0.11$ (red circles), $0.12$ (gray squares), and $0.16$ (blue diamonds). The dashed lines guide the eye. 
\label{fig：1fnoise}}
 \end{figure}

Resistance fluctuations $\Delta R(t)=R(t)-\langle R\rangle$, where $\langle R\rangle$ is the time-averaged resistance and $t$ is time, were measured at $3~\mathrm {K}<T<280~\mathrm {K}$~\cite{supp}. 
While RTN appeared intermittently near $T=T_{\mathrm{max}}$ for 
$x=0.11$ and $0.12$, the noise with power spectral density $S_R\propto 1/f^\alpha$ and $\alpha=1.0\pm0.3$, was observed in 
all samples over the entire $T$ range. Figure~\ref{fig：1fnoise} shows that the more insulating samples (lower $x$) exhibit higher noise at a given $T/T_\mathrm{max}$, and the same sample exhibits higher noise in the phase-coexistence region than in the insulating phase. The behavior of the $1/f$ noise is discussed in more detail in the Supplemental Material~\cite{supp}.
\begin{figure*}
 \includegraphics[width=2\columnwidth]{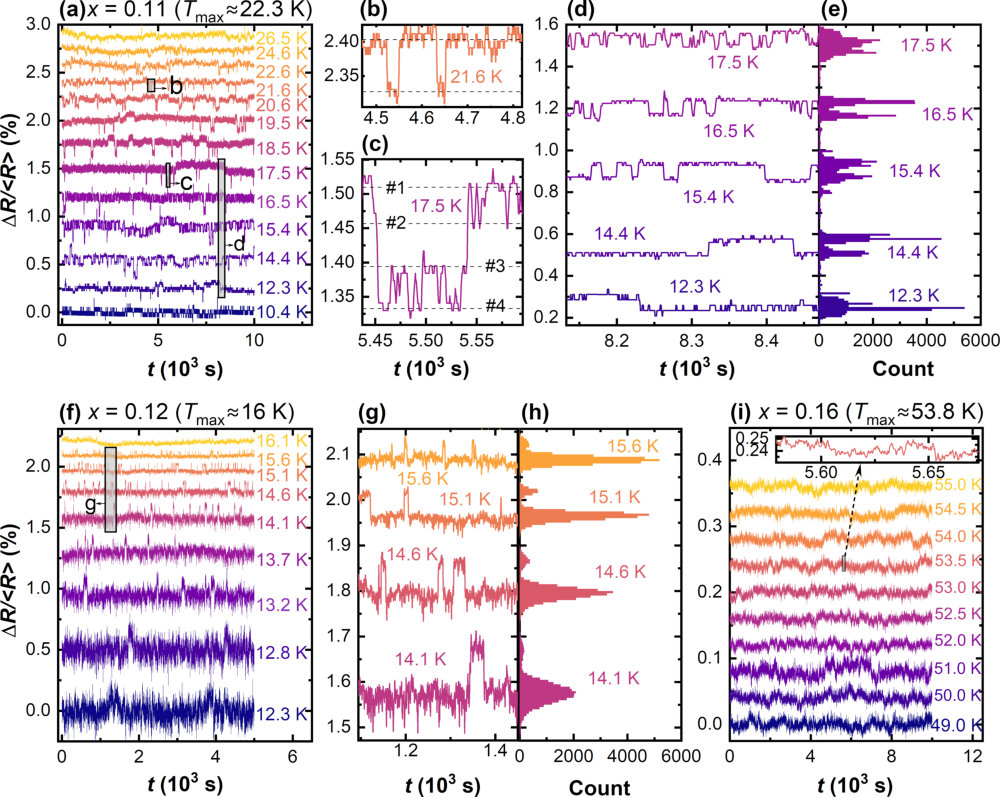}
\caption{
$\Delta R/\langle R \rangle$ vs time 
for (a)--(d) $x=0.11$, (f)--(g) $x=0.12$, (i) $x=0.16$ at various $T$ near their $T_\mathrm{max}$. 
Well-defined TLS is observed from $x=0.11$ and $0.12$, as demonstrated by the zoomed-in time-domain traces in (b)--(d) and (g), and by the multiple peaks in the  
histograms of $\Delta R/\langle R \rangle$ in (e) and (h) [see Supplemental Material, Fig.~S2~\cite{supp}, for histograms of all 
data in (a) and (b)]. Gray rectangles denote regions enlarged in other panels or the inset. 
All traces are vertically shifted by an arbitrary amount for clarity. 
\label{Fig:time_domain_RTN}}
 \end{figure*}
Here we focus on RTN, which was observed in the $T$ ranges highlighted in Fig.~\ref{Fig:intro}(c). For example, for $x=0.11$ [Fig.~\ref{Fig:time_domain_RTN}(a)], some discrete resistance jumps appear at a few kelvins above $T_\mathrm{max}$, followed by the emergence of distinct TLS, at $21.6~\mathrm{K}$ [Fig.~\ref{Fig:time_domain_RTN}(b)] and below. Upon further cooling, another, ``faster'' TLS event appears between $17.5~\mathrm {K}$ and $12.3~\mathrm {K}$, resulting in four discrete resistance states of the sample [Fig.~\ref{Fig:time_domain_RTN}(c)]. The TLS events occur less frequently at lower $T$ [Fig.~\ref{Fig:time_domain_RTN}(d)], 
and almost disappear below $12.3~\mathrm {K}$. The emergence of discrete states and their evolution with $T$ can also be seen from the peaks in the histograms of resistance fluctuations [Fig.~\ref{Fig:time_domain_RTN}(e)].
Similar RTN has been observed from the $x=0.12$ sample [Figs.~\ref{Fig:time_domain_RTN}(f)--(h)]: well-defined TLS appears at $15.6~\mathrm{K}<T_\mathrm{max}\approx16~\mathrm{K}$, and its occurrence rate decreases on cooling. 

To test the effect of thermal cycling on RTN, we performed the following protocols. i) We heated the samples only up to 
$10~\mathrm{K}$ above their $T_\mathrm{max}$, and found that the RTN with the same statistical features, discussed below, reappeared at the same $T$. ii) In contrast, by warming up to $50~\mathrm{K}$ above $T_\mathrm{max}$, we found that the RTN could assume quantitatively different statistics or even disappear. To illustrate, ten consecutive noise measurements were performed on the $x=0.11$ sample at $21.5~\mathrm{K}\lesssim~T_\mathrm{max}\approx22.3~\mathrm{K}$ following the identical thermal-cycling protocol between $21.5~\mathrm{K}$ and $70~\mathrm{K}$. Well-defined RTN was observed in three of the ten runs (See Supplemental Material, Fig.~S3~\cite{supp}). This stochastic reappearance of RTN after large thermal cycles is reminiscent of the similar phenomenon in manganites~\cite{Raquet_Anane_PRL2000}. In contrast, well-defined RTN was never observed from the $x=0.16$ sample near its $T_\mathrm{max}$ [Fig.~\ref{Fig:time_domain_RTN}(i)], despite many repeated measurements following small or large thermal cycles (See Supplemental Material, Fig.~S4~\cite{supp}).

\begin{figure}
 \includegraphics[width=1\columnwidth]{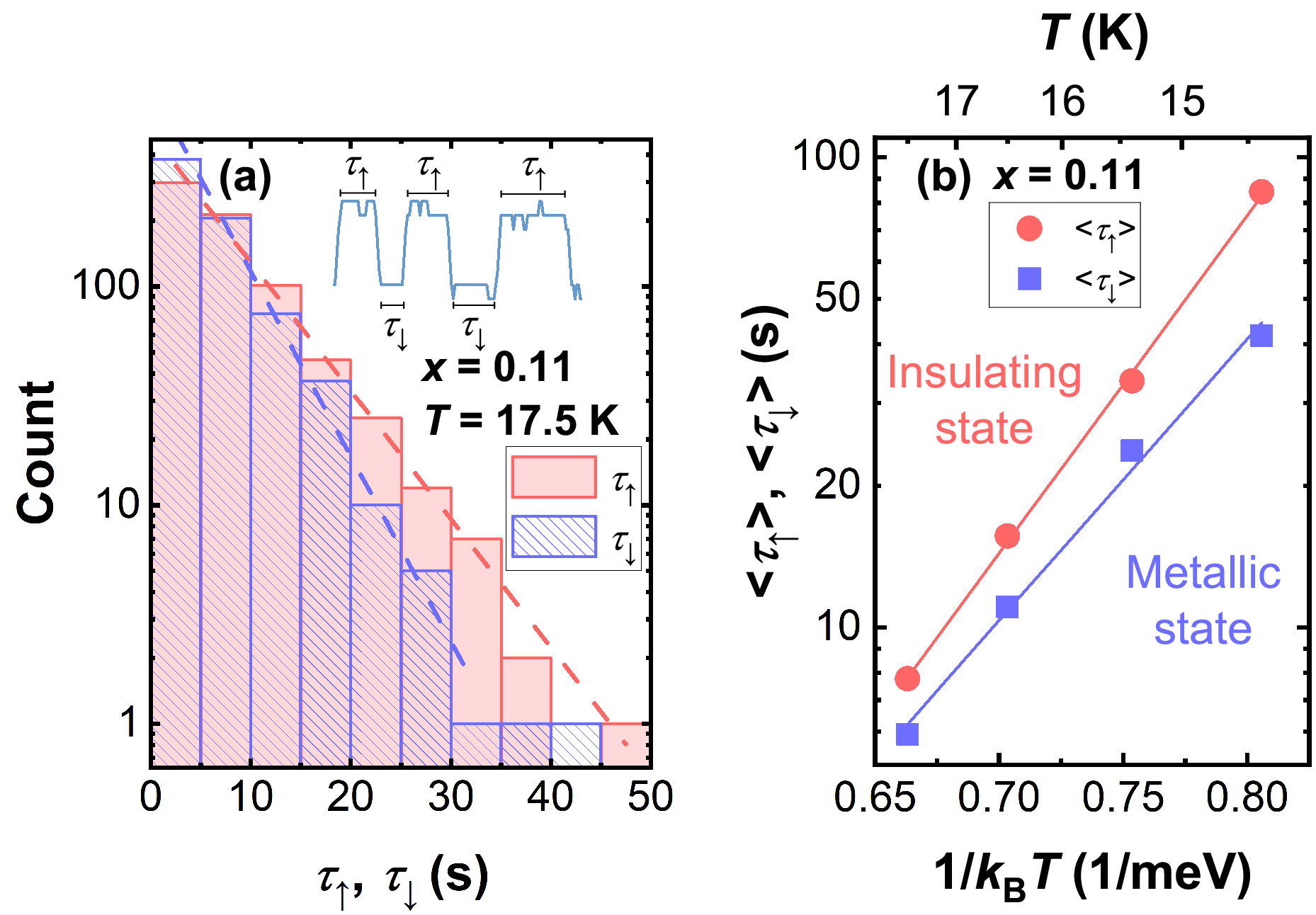}
\caption{
(a) Distributions of dwell times $\tau_\uparrow$ and $\tau_\downarrow$ derived from the ``faster TLS'' in $x=0.11$ at $T=17.5~\mathrm {K}$ [Fig.~\ref{Fig:time_domain_RTN}(a)].  The dashed lines are exponential fits.
The sketch of RTN demonstrates the determination of $\tau_\uparrow$ and $\tau_\downarrow$. (b) Arrhenius plot of $\langle \tau_\uparrow \rangle$ and $\langle \tau_\downarrow \rangle$ derived from the ``faster TLS'' in the $x=0.11$ data in Fig.~\ref{Fig:time_domain_RTN}(a). The 
data (symbols) are well described by 
$\langle \tau_i \rangle=\tau_{0,i}\exp\left( {{E_i}/{k_\mathrm{B}T}}\right)$,
as indicated by the solid fitting lines, suggesting $\tau_{0,\uparrow}=0.13\pm0.03~\mathrm{ms}$, $\tau_{0,\downarrow}=0.66\pm0.21~\mathrm{ms}$, $E_{\uparrow}=16.6\pm0.4~\mathrm{meV}$, and $E_{\downarrow}=13.8\pm0.8~\mathrm{meV}$. The uncertainties are determined from the standard deviation of the fits. 
\label{Fig:statistics}}
 \end{figure}

To understand its physical meaning, we perform statistical analysis on the RTN near $T_\mathrm{max}$. 
In particular, we focus on noise in $x=0.11$ at $T=17.5~\mathrm{K}$, which shows four states [marked as states $\#1$--$\#4$ in Fig~\ref{Fig:time_domain_RTN}(c)] arising from the superposition of two TLS systems: one switching between states $\#1$ and $\#2$ (or states $\#3$ and $\#4$), and the other switching between states $\#1$ and $\#3$ (or states $\#2$ and $\#4$). 
The former has many more switching events to provide accurate statistics, and hence we refer to it as the ``faster TLS''.  We analyze its statistics by considering states $\#1$ and $\#3$ as the higher resistance state (denoted by $\uparrow$), and states $\#2$ and $\#4$ as the lower resistance state (denoted by $\downarrow$). (Analyzing the other TLS system in $x=0.11$ or that in $x=0.12$ yields qualitatively consistent statistics; see Supplemental Material, Fig.~S5~\cite{supp}.) The histograms of the dwell times 
$\tau_\uparrow$ and $\tau_\downarrow$ reveal an exponential distribution $P(\tau_i)=({1}/{\langle \tau_i \rangle})\exp\left(-{{\tau_i}/{\langle \tau_i \rangle}}\right)$,
where $P$ denotes the probability density, $i=\uparrow, \downarrow$, and $\langle \tau_i \rangle$ is the mean dwell time [Fig.~\ref{Fig:statistics}(a)]. From a statistical perspective, this exponential distribution immediately suggests that the RTN arises from a memoryless process~\cite{ROSS_ED2014}. In addition, due to the presence of discrete states, the resistance noise exhibits a non-Gaussian amplitude distribution [Figs.~\ref{Fig:time_domain_RTN}(e) and (h)]. While non-Gaussian amplitude distributions can sometimes indicate a nonstationary process, we have analyzed higher-order statistics~\cite{Seidler_Solin_PRB1996} of the noise and found no signature of nonstationarity (see Supplemental Material, Fig.~S6~\cite{supp}).

The TLS occurs less frequently at lower $T$, which can be quantitatively characterized by the $T$-dependence of the mean dwell time $\langle \tau_i \rangle$. Taking the ``faster TLS" of $x=0.11$ at $T=14.4$--$17.5 ~\mathrm{K}$ as an example, Fig.~\ref{Fig:statistics}(b) shows that, although limited in a $T$ range, $\langle \tau_i \rangle$ decreases with $T$ in a thermally activated manner, $\langle \tau_i \rangle=\tau_{0,i}\exp\left( {{E_i}/{k_\mathrm{B}T}}\right)$,
where $\tau_{0,i}$ and $E_i$ are the attempt times and energy barriers, respectively.  The observed behavior is consistent with an asymmetrical TLS system \cite{Weissman_RVP1988, Raquet_Anane_PRL2000} where an energy barrier separates the two states with height $E_i$ from each side [Fig.~\ref{Fig:intro}(a)], as expected for 
a domain fluctuating between two coexisting phases.

The RTN observed near $T_\mathrm{max}$ is, therefore, attributed to domains that switch back and forth  
between metallic and insulating states, driven by thermal fluctuations.  Our results indicate that such domains are located within the dominant conduction channel [Fig.~\ref{Fig:intro}(a)], which forms the metallic percolating cluster once $T$ decreases below $T_\mathrm{max}$. In other words, we find that the dynamics near the resistivity maximum is consistent with the percolation picture~\cite{Tan_Dobrosavljevic_Crystals2022}. This interpretation also consistently explains other aspects of our observations. The fact that RTN can change its statistics or disappear after large thermal cycles above $T_\mathrm{max}$ can be understood in terms of the random formation of fluctuating domains and of a percolating cluster when cooling from temperatures 
far above the phase coexistence region [cf. Fig.~\ref{Fig:intro}(b)]. If no fluctuating domains form within the percolating cluster, the RTN disappears; if multiple fluctuating domains are present, their RTN signals superpose.
Notably, from the measurements where RTN is observed, the barrier heights $E_i$ are always found to be 
$\sim10~\mathrm{meV}$, with the magnitude of RTN remaining 
$\sim0.1\%$ of the total resistance. These approximately constant 
energy and fluctuation scales possibly suggest a steady characteristic volume  
of the fluctuating domains. Finally, a domain fluctuating between the two phases is a simple two-level system at thermal equilibrium, in agreement with the stationary and memoryless nature of RTN. 

The absence of the distinct RTN in 
the $x=0.16$ sample, which is also in the phase-coexistence region~\cite{Pustogow_Rosslhuber_npjqm2021}, can be ascribed to at least two possible reasons. First, since $x=0.16$ is closer to the metallic phase, its $T_\mathrm{max}$ might be affected by both the percolation and the thermal formation/destruction of Landau quasiparticles, in contrast to $x=0.11$ and $x=0.12$ where the percolation effect dominates. Second, $T_\mathrm{max}$ of this sample is so high that the resistivity values of the metallic and insulating phases are comparable, and thus the domain fluctuations cannot give rise to well-resolved TLS in the global resistance. 

Noise near the Mott MIT has also been investigated in other organic and inorganic systems, where slow dynamics have been reported and attributed to various mechanisms. In deuterated \Br~\cite{Brandenburg_Muller_NJP2012, Hartmann_Zielke_PRL2015}, resistance noise suddenly slows down in the vicinity of the critical point of the Mott MIT, where the transition becomes continuous. The slowing down was naturally attributed to the criticality of the Mott transition and was believed to be universal near a Mott critical point. However, this interpretation has been challenged by a very recent work~\cite{Thyzel_Franke_PRB2025}, which finds such slow dynamics absent in pressurized \Cl. It is worth noting that in those two Mott organic systems, a glass-like freezing of the BEDT-TTF molecules' terminal ethylene groups also contributes to the slow resistance noise~\cite{Muller_Thomas_Crystals2018}, which can complicate the interpretation of the noise results. In the meantime, glassy electron kinetics probed near the first-order Mott MIT of Ruddlesden-Popper ruthenate  Ca$_{3}$(Ru$_{0.9}$Ti$_{0.1}$)$_{2}$O$_{7}$ were ascribed to accompanying lattice distortion and magnetic defects~\cite{Kumbhakar_Islam_PRB2022}. By comparison, \stf~possesses genuine Mott transition without interference from additional competing degrees of freedom. Our observation of stationary TLS thus points to an intrinsic feature of the Mott MIT, in line with the theoretical prediction that Mott localization tends to suppress the glassy behavior of electrons~\cite{Dobrosavljevic_Tanaskovic_PRL2003}.

By combining transport and noise spectroscopy across the Mott transition in \stf, we identify time-domain signatures of dynamical phase coexistence. Our results demonstrate that strong electronic correlations, rather than static disorder, control the arrest of charge motion at the genuine Mott transition. We establish that the resistivity maximum at $T = T_\mathrm{max}$ in bandwidth-tuned Mott systems marks a dynamical boundary reflecting real-time competition between
metallic and insulating tendencies, rather than a mere crossover scale. Time-resolved transport reveals that conduction in this regime is governed by intermittency, rather than by a smooth, homogeneous loss of coherence. More broadly, our work elevates time-resolved transport to a powerful probe for identifying Mott phase competition and provides a unified framework for interpreting the resistivity maxima in correlated materials. 

\textit{Acknowledgment}---We thank G. Untereiner for assistance with sample preparation. This work was supported by NSF DMR-2104193 (D.P.), NSF DMR-2409911 (V.D.), and the National High Magnetic Field Laboratory through NSF Cooperative Agreement No. DMR-2128556, and the State of Florida. We thank the Deutsche Forschungsgemeinschaft (DFG) for funding via DR 228/74-1.

\bibliography{Reference}

@article{Raquet_Anane_PRL2000,
  title = {Noise Probe of the Dynamic Phase Separation in {${\mathrm{La}}_{2/3}{\mathrm{Ca}}_{1/3}{\mathrm{MnO}}_{3}$}},
  author = {Raquet, B. and Anane, A. and Wirth, S. and Xiong, P. and von Moln\'ar, S.},
  journal = {Phys. Rev. Lett.},
  volume = {84},
  issue = {19},
  pages = {4485--4488},
  numpages = {0},
  year = {2000},
  month = {May},
  publisher = {American Physical Society},
  doi = {10.1103/PhysRevLett.84.4485},
  url = {https://link.aps.org/doi/10.1103/PhysRevLett.84.4485}
}

@article{Weissman_RVP1988,
  title = {$\frac{1}{f}$ noise and other slow, nonexponential kinetics in condensed matter},
  author = {Weissman, M. B.},
  journal = {Rev. Mod. Phys.},
  volume = {60},
  issue = {2},
  pages = {537--571},
  numpages = {0},
  year = {1988},
  month = {Apr},
  publisher = {American Physical Society},
  doi = {10.1103/RevModPhys.60.537},
  url = {https://link.aps.org/doi/10.1103/RevModPhys.60.537}
}

@incollection{ROSS_ED2014,
title = {The Exponential Distribution and the {Poisson} Process},
booktitle = {Introduction to Probability Models},
publisher = {Academic Press},
edition = {11th},
address = {Boston},
pages = {277-356},
year = {2014},
isbn = {978-0-12-407948-9},
doi = {https://doi.org/10.1016/B978-0-12-407948-9.00005-0},
url = {https://www.sciencedirect.com/science/article/pii/B9780124079489000050},
author = {Sheldon Ross}
}

@article{Pustogow_Rosslhuber_npjqm2021,
author={Pustogow, A.
and R{\"o}sslhuber, R.
and Tan, Y.
and Uykur, E.
and B{\"o}hme, A.
and Wenzel, M.
and Saito, Y.
and L{\"o}hle, A.
and H{\"u}bner, R.
and Kawamoto, A.
and Schlueter, J. A.
and Dobrosavljevi{\'{c}}, V.
and Dressel, M.},
title={Low-temperature dielectric anomaly arising from electronic phase separation at the {Mott} insulator-metal transition},
journal={npj Quantum Mater.},
year={2021},
month={Jan},
day={27},
volume={6},
number={1},
pages={9},
url={https://doi.org/10.1038/s41535-020-00307-0}
}

@article{Pinteric_Culo_PRB2014,
  title = {Anisotropic charge dynamics in the quantum spin-liquid candidate {$\kappa$-(BEDT\--TTF)$_2$\-Cu$_2$(CN)$_3$}},
  author = {Pinteri\ifmmode \acute{c}\else \'{c}\fi{}, M. and \ifmmode \check{C}\else \v{C}\fi{}ulo, M. and Milat, O. and Basleti\ifmmode \acute{c}\else \'{c}\fi{}, M. and Korin-Hamzi\ifmmode \acute{c}\else \'{c}\fi{}, B. and Tafra, E. and Hamzi\ifmmode \acute{c}\else \'{c}\fi{}, A. and Ivek, T. and Peterseim, T. and Miyagawa, K. and Kanoda, K. and Schlueter, J. A. and Dressel, M. and Tomi\ifmmode \acute{c}\else \'{c}\fi{}, S.},
  journal = {Phys. Rev. B},
  volume = {90},
  issue = {19},
  pages = {195139},
  numpages = {13},
  year = {2014},
  month = {Nov},
  publisher = {American Physical Society},
  doi = {10.1103/PhysRevB.90.195139},
  url = {https://link.aps.org/doi/10.1103/PhysRevB.90.195139}
}

@article{Brandenburg_Muller_NJP2012,
doi = {10.1088/1367-2630/14/2/023033},
url = {https://doi.org/10.1088/1367-2630/14/2/023033},
year = {2012},
month = {feb},
publisher = {IOP Publishing},
volume = {14},
number = {2},
pages = {023033},
author = {Brandenburg, Jens and Müller, Jens and Schlueter, John A},
title = {Sudden slowing down of charge carrier dynamics at the {Mott} metal–insulator transition in {$\kappa$-(D$_8$-BEDT-TTF)$_2$Cu[N(CN)$_2$]Br}},
journal = {New J. Phys.},
}

@article{Hartmann_Zielke_PRL2015,
  title = {Critical Slowing Down of the Charge Carrier Dynamics at the {Mott} Metal-Insulator Transition},
  author = {Hartmann, Benedikt and Zielke, David and Polzin, Jana and Sasaki, Takahiko and M\"uller, Jens},
  journal = {Phys. Rev. Lett.},
  volume = {114},
  issue = {21},
  pages = {216403},
  numpages = {5},
  year = {2015},
  month = {May},
  publisher = {American Physical Society},
  doi = {10.1103/PhysRevLett.114.216403},
  url = {https://link.aps.org/doi/10.1103/PhysRevLett.114.216403}
}

@article{Thyzel_Franke_PRB2025,
  title = {Absence of critical fluctuations at the bandwidth-controlled {Mott} transition in a molecular conductor},
  author = {Thyzel, Tim and Franke, Lars and Garst, Markus and Schubert, Harald and Lang, Michael and Sasaki, Takahiko and M\"uller, Jens},
  journal = {Phys. Rev. B},
  volume = {112},
  issue = {12},
  pages = {L121101},
  numpages = {7},
  year = {2025},
  month = {Sep},
  publisher = {American Physical Society},
  doi = {10.1103/j1jp-yy68},
  url = {https://link.aps.org/doi/10.1103/j1jp-yy68}
}

@article{Muller_Thomas_Crystals2018,
author = {M\"uller, Jens and Thomas, Tatjana},
TITLE = {Low-Frequency Dynamics of Strongly Correlated Electrons in {(BEDT-TTF)$_2$X} Studied by Fluctuation Spectroscopy},
JOURNAL = {Crystals},
VOLUME = {8},
YEAR = {2018},
issue = {4},
pages = {166},
URL = {https://www.mdpi.com/2073-4352/8/4/166},
ISSN = {2073-4352},
DOI = {10.3390/cryst8040166},
}

@article{Thomas_Agarmani_npjs2024,
author={Thomas, Tatjana
and Agarmani, Yassine
and Hartmann, Steffi
and Kartsovnik, Mark
and Kushch, Natalia
and Winter, Stephen M.
and Schmid, Sebastian
and Lunkenheimer, Peter
and Lang, Michael
and M{\"u}ller, Jens},
title={Slow and non-equilibrium dynamics due to electronic ferroelectricity in a strongly-correlated molecular conductor},
journal={npj Spintron.},
year={2024},
month={Jun},
day={28},
volume={2},
number={1},
pages={24},
issn={2948-2119},
doi={10.1038/s44306-024-00022-7},
url={https://doi.org/10.1038/s44306-024-00022-7}
}

@article{Kumbhakar_Islam_PRB2022,
  title = {Glassy electrons at the first-order {Mott} metal-insulator transition},
  author = {Kumbhakar, Shreya and Islam, Saurav and Mao, Zhiqiang and Wang, Yu and Ghosh, Arindam},
  journal = {Phys. Rev. B},
  volume = {106},
  issue = {20},
  pages = {L201112},
  numpages = {6},
  year = {2022},
  month = {Nov},
  publisher = {American Physical Society},
  doi = {10.1103/PhysRevB.106.L201112},
  url = {https://link.aps.org/doi/10.1103/PhysRevB.106.L201112}
}

@article{Dobrosavljevic_Tanaskovic_PRL2003,
  title = {Glassy Behavior of Electrons Near Metal-Insulator Transitions},
  author = {Dobrosavljevi\ifmmode \acute{c}\else \'{c}\fi{}, V. and Tanaskovi\ifmmode \acute{c}\else \'{c}\fi{}, D. and Pastor, A. A.},
  journal = {Phys. Rev. Lett.},
  volume = {90},
  issue = {1},
  pages = {016402},
  numpages = {4},
  year = {2003},
  month = {Jan},
  publisher = {American Physical Society},
  doi = {10.1103/PhysRevLett.90.016402},
  url = {https://link.aps.org/doi/10.1103/PhysRevLett.90.016402}
}

@article{Saito_Lohle_Crystals2021,
AUTHOR = {Saito, Yohei and L{\"o}hle, Anja and Kawamoto, Atsushi and Pustogow, Andrej and Dressel, Martin},
TITLE = {Pressure-Tuned Superconducting Dome in Chemically-Substituted {$\kappa$-(BEDT\--TTF)$_2$\-Cu$_2$(CN)$_3$}},
JOURNAL = {Crystals},
VOLUME = {11},
YEAR = {2021},
issue = {7},
pages = {817},
URL = {https://www.mdpi.com/2073-4352/11/7/817},
ISSN = {2073-4352},
}

@article{Saito_Rosslhuber_JMCC2021,
author ={Saito, Yohei and R{\"o}sslhuber, Roland and L{\"o}hle, Anja and Sanz Alonso, Miriam and Wenzel, Maxim and Kawamoto, Atsushi and Pustogow, Andrej and Dressel, Martin},
title  ={Chemical tuning of molecular quantum materials {$\kappa$-[(BEDT\--TTF)$_{1-x}$\-(BEDT\--STF)$_{x}$]$_2$\-Cu$_2$(CN)$_3$}: from the {Mott}-insulating quantum spin liquid to metallic {Fermi} liquid},
journal  ={J. Mater. Chem. C},
year  ={2021},
volume  ={9},
issue  ={33},
pages  ={10841-10850},
publisher  ={The Royal Society of Chemistry},
doi  ={10.1039/D1TC00785H},
}

@misc{supp, 
note = "See Supplemental Material at [URL] for the experimental details, the $1/f$ noise, and additional analyses on RTN, which includes Refs.~[32-34]",}

@article{Seidler_Solin_PRB1996,
  title = {Non-Gaussian $1/f$ noise: Experimental optimization and separation of high-order amplitude and phase correlations},
  author = {Seidler, G. T. and Solin, S. A.},
  journal = {Phys. Rev. B},
  volume = {53},
  issue = {15},
  pages = {9753--9759},
  numpages = {0},
  year = {1996},
  month = {Apr},
  publisher = {American Physical Society},
  doi = {10.1103/PhysRevB.53.9753},
  url = {https://link.aps.org/doi/10.1103/PhysRevB.53.9753}
}

@misc{Wang_Dobrosavljevic_AXV2025,
      title={Failed superconductivity in a {Mott} spin liquid material}, 
      author={Yuxin Wang and Vladimir Dobrosavljevi\'{c} and Eun Sang Choi and Yohei Saito and Atsushi Kawamoto and Andrej Pustogow and Martin Dressel and Dragana Popovi\'{c}},
      year={2025},
      eprint={2507.10832},
      archivePrefix={arXiv},
      url={https://arxiv.org/abs/2507.10832}, 
}

@Article{Pustogow_Saito_NatComm2021,
author={Pustogow, Andrej
and Saito, Yohei
and L{\"o}hle, Anja
and Sanz Alonso, Miriam
and Kawamoto, Atsushi
and Dobrosavljevi{\'{c}}, Vladimir
and Dressel, Martin
and Fratini, Simone},
title={Rise and fall of {Landau}'s quasiparticles while approaching the {Mott} transition},
journal={Nat. Commun.},
year={2021},
month={Mar},
day={10},
volume={12},
number={1},
pages={1571},
issn={2041-1723},
doi={10.1038/s41467-021-21741-z},
url={https://doi.org/10.1038/s41467-021-21741-z}
}

@book{Mott_MIT1990,
  title={Metal-Insulator Transitions},
  author={Mott, N.},
  isbn={9781466576452},
  year={1990},
  publisher={Taylor \& Francis}
}

@Article{Tan_Dobrosavljevic_Crystals2022,
AUTHOR = {Tan, Yuting and Dobrosavljevi\'{c}, Vladimir and Rademaker, Louk},
TITLE = {How to Recognize the Universal Aspects of {Mott} Criticality?},
JOURNAL = {Crystals},
VOLUME = {12},
YEAR = {2022},
ISSUE = {7},
pages = {932},
ISSN = {2073-4352},
DOI = {10.3390/cryst12070932}
}

@article{Dressel_Tomic_AP2020,
author = {Martin Dressel and Silvia Tomi\'{c}},
title = {Molecular quantum materials: electronic phases and charge dynamics in two-dimensional organic solids},
journal = {Adv. Phys.},
volume = {69},
number = {1}, 
pages = {1--120},
year = {2020},
publisher = {Taylor \& Francis},
doi = {10.1080/00018732.2020.1837833},
URL = {https://doi.org/10.1080/00018732.2020.1837833}
}

@article{Radonjic_Tanaskovic_PRB2012,
  title = {{Wigner-Mott} scaling of transport near the two-dimensional metal-insulator transition},
  author = {Radonji\ifmmode \acute{c}\else \'{c}\fi{}, M. M. and Tanaskovi\ifmmode \acute{c}\else \'{c}\fi{}, D. and Dobrosavljevi\ifmmode \acute{c}\else \'{c}\fi{}, V. and Haule, K. and Kotliar, G.},
  journal = {Phys. Rev. B},
  volume = {85},
  issue = {8},
  pages = {085133},
  numpages = {7},
  year = {2012},
  month = {Feb},
  publisher = {American Physical Society},
  doi = {10.1103/PhysRevB.85.085133},
  url = {https://link.aps.org/doi/10.1103/PhysRevB.85.085133}
}

@article{Moon_Han_PRB2020,
  title = {Quantum critical scaling for finite-temperature {Mott}-like metal-insulator crossover in few-layered {${\mathrm{MoS}}_{2}$}},
  author = {Moon, Byoung Hee and Han, Gang Hee and Radonji\ifmmode \acute{c}\else \'{c}\fi{}, Milo\ifmmode \check{s}\else \v{s}\fi{} M. and Ji, Hyunjin and Dobrosavljevi\ifmmode \acute{c}\else \'{c}\fi{}, Vladimir},
  journal = {Phys. Rev. B},
  volume = {102},
  issue = {24},
  pages = {245424},
  numpages = {9},
  year = {2020},
  month = {Dec},
  publisher = {American Physical Society},
  doi = {10.1103/PhysRevB.102.245424},
  url = {https://link.aps.org/doi/10.1103/PhysRevB.102.245424}
}

@article{Shashkin_Melnikov_PRB2020,
  title = {Manifestation of strong correlations in transport in ultraclean {SiGe/Si/SiGe} quantum wells},
  author = {Shashkin, A. A. and Melnikov, M. Yu. and Dolgopolov, V. T. and Radonji\ifmmode \acute{c}\else \'{c}\fi{}, M. M. and Dobrosavljevi\ifmmode \acute{c}\else \'{c}\fi{}, V. and Huang, S.-H. and Liu, C. W. and Zhu, Amy Y. X. and Kravchenko, S. V.},
  journal = {Phys. Rev. B},
  volume = {102},
  issue = {8},
  pages = {081119},
  numpages = {5},
  year = {2020},
  month = {Aug},
  publisher = {American Physical Society},
  doi = {10.1103/PhysRevB.102.081119},
  url = {https://link.aps.org/doi/10.1103/PhysRevB.102.081119}
}

@Article{Pustogow_Bories_NatMater2018,
author={Pustogow, A.
and Bories, M.
and L{\"o}hle, A.
and R{\"o}sslhuber, R.
and Zhukova, E.
and Gorshunov, B.
and Tomi{\'{c}}, S.
and Schlueter, J. A.
and H{\"u}bner, R.
and Hiramatsu, T.
and Yoshida, Y.
and Saito, G.
and Kato, R.
and Lee, T.-H.
and Dobrosavljevi{\'{c}}, V.
and Fratini, S.
and Dressel, M.},
title={Quantum spin liquids unveil the genuine {Mott} state},
journal={Nat. Mater.},
year={2018},
month={Sep},
day={01},
volume={17},
number={9},
pages={773-777},
issn={1476-4660},
doi={10.1038/s41563-018-0140-3},
url={https://doi.org/10.1038/s41563-018-0140-3}
}

@Article{Li_Jiang_Nature2021,
author={Li, Tingxin
and Jiang, Shengwei
and Li, Lizhong
and Zhang, Yang
and Kang, Kaifei
and Zhu, Jiacheng
and Watanabe, Kenji
and Taniguchi, Takashi
and Chowdhury, Debanjan
and Fu, Liang
and Shan, Jie
and Mak, Kin Fai},
title={Continuous {Mott} transition in semiconductor moir{\'e} superlattices},
journal={Nature},
year={2021},
month={Sep},
day={01},
volume={597},
number={7876},
pages={350-354},
issn={1476-4687},
doi={10.1038/s41586-021-03853-0},
url={https://doi.org/10.1038/s41586-021-03853-0}
}

@article{spivakRMP,
  title = {Colloquium: Transport in strongly correlated two dimensional electron fluids},
  author = {Spivak, B. and Kravchenko, S. V. and Kivelson, S. A. and Gao, X. P. A.},
  journal = {Rev. Mod. Phys.},
  volume = {82},
  issue = {2},
  pages = {1743--1766},
  numpages = {0},
  year = {2010},
  month = {May},
  publisher = {American Physical Society},
  doi = {10.1103/RevModPhys.82.1743},
  url = {https://link.aps.org/doi/10.1103/RevModPhys.82.1743}
}

@article{finkelstein,
  title={Flow diagram of the metal--insulator transition in two dimensions},
  author={Anissimova, S and Kravchenko, SV and Punnoose, A and Finkel’Stein, AM and Klapwijk, TM},
  journal={Nat. Phys.},
  volume={3},
  number={10},
  pages={707--710},
  year={2007},
  publisher={Nature Publishing Group UK London}
}

@article{wigner-mott,
  title={Coulomb correlations and the {Wigner--Mott} transition},
  author={Camjayi, A and Haule, K and Dobrosavljevi{\'c}, V and Kotliar, G},
  journal={Nat. Phys.},
  volume={4},
  number={12},
  pages={932--935},
  year={2008},
  publisher={Nature Publishing Group UK London}
}

@article{manganitesRMP2001,
  title = {The physics of manganites: Structure and transport},
  author = {Salamon, Myron B. and Jaime, Marcelo},
  journal = {Rev. Mod. Phys.},
  volume = {73},
  issue = {3},
  pages = {583--628},
  numpages = {0},
  year = {2001},
  month = {Aug},
  publisher = {American Physical Society},
  doi = {10.1103/RevModPhys.73.583},
  url = {https://link.aps.org/doi/10.1103/RevModPhys.73.583}
}

@article{heavyfermion1984RMP,
  title = {Heavy-fermion systems},
  author = {Stewart, G. R.},
  journal = {Rev. Mod. Phys.},
  volume = {56},
  issue = {4},
  pages = {755--787},
  numpages = {0},
  year = {1984},
  month = {Oct},
  publisher = {American Physical Society},
  doi = {10.1103/RevModPhys.56.755},
  url = {https://link.aps.org/doi/10.1103/RevModPhys.56.755}
}

\clearpage

\clearpage
\setcounter{figure}{0}
\makeatletter
\makeatletter \renewcommand{\fnum@figure}{{\figurename~S\thefigure}}
\makeatother

\onecolumngrid
\begin{center}
\textbf{\Large Supplemental Material for}\\
\vspace*{6pt}
\textbf{\Large Mott Intermittency at the Metal–Insulator Boundary}\\
\vspace{1em}
\end{center}
\twocolumngrid

\section{Experimental details\label{Experimental details}}
The materials studied here are plate-like single crystals of \STF~(abbreviated~\stf)~with $x=0.11$ ($a \times b \times c=0.13 \times 0.7 \times 1.3 ~\textrm{mm}^3$), $x=0.12$ ($0.04 \times 0.4 \times 0.9~\textrm{mm}^3$), and $x=0.16$ ($0.04 \times 1.0 \times 1.4~\textrm{mm}^3$) prepared by electrochemical oxidation~[S1]. Here, BEDT\--TTF and BEDT\--STF stand for bis\-(ethylenedithio)\-tetrathiafulvalene and bis\-(ethylenedithio)\-diselenium\-dithiafulvalene, respectively. From BEDT\--TTF to BEDT\--STF, two S atoms in the inner ring are replaced by Se that have more extended orbitals. Such a substitution thus increases the bandwidth and tunes the system from a Mott insulator towards a Fermi liquid. Along the $c$-axis (the most conductive axis~[S2]) of these samples, electrical contacts were made by the DOTITE XC-12 carbon paste connected to $25~\mu$m-thick gold wires, with a typical distance of $\sim 0.4~\mathrm{mm}$ between the voltage contacts. 

The experiments were conducted in $^4$He cryostats over the temperature ($T$) range $3–280~\mathrm{K}$ for the $x=0.11$ and $0.12$ samples, and $5–200~\mathrm{K}$ for the $x=0.16$ sample.  Transport and noise were measured with a Lakeshore 372 AC resistance bridge or Signal Recovery 7265 lock-in amplifier and LI-75A voltage preamplifier using the standard four-probe ac technique (typically $\sim 17~\mathrm{Hz}$), except that the noise data on $x=0.11$ were acquired using a five-probe setup~[S3] to minimize the effects of $T$ fluctuations, contact noise, and excitation source fluctuations.
Both the resistance bridge and the lock-in amplifier measured the in-phase and out-of-phase components of the signal simultaneously. 
The power spectrum of the out-of-phase component, which contains only the 
background noise 
including instrumental and environmental contributions, 
was consistently several orders of magnitude lower than the power spectrum of the in-phase component. When RTN was observed in the in-phase component, 
the out-of-phase component did not show TLS. Since the in-phase component of the signal contains both sample resistance noise and background, 
the background contribution was nevertheless subtracted from the power spectrum of the in-phase component 
to obtain the resistance noise power spectrum. 
To avoid aliasing, time constants of the instruments were set to $0.5$~s or $1~\mathrm{s}$, while the data were collected at a time interval of $0.2$~s or $0.5~\mathrm{s}$. 
At each measured $T$, excitation currents ranging from $50~\mathrm{nA}$ to $100~\mathrm{\mu A}$ were employed for the noise measurements and yielded quantitatively consistent results. This consistency ruled out heating effects and contact-noise contributions, as both are expected to depend on the excitation current. 

In parallel with transport and noise measurements, sample temperatures were monitored with a Lakeshore 336 temperature controller and a Cernox thermometer placed a few mm below the sample.
Typically, the standard deviation of $T$ fluctuations was about $1~\mathrm{mK}$ for $T\leq100~\mathrm {K}$, and a few $\mathrm{mK}$ for $T>100~\mathrm {K}$. Temperature never exhibited any two-level switching, and 
thus it did not account for the TLS observed in the  
resistance of the samples. To quantify the 
effect of $T$ fluctuations on the measured  
resistance fluctuations, we calculated the power spectral density 
$S_{T}$ of $T$ fluctuations and the resulting
resistance fluctuations $S_{T\rightarrow R}=S_{T}(d\langle R\rangle/dT)^2$.
By comparing $S_{T\rightarrow R}$ and the measured power spectral density of the resistance fluctuations $S_{R}$, we found $S_{T\rightarrow R}$ to be at least one order of magnitude lower than $S_{R}$ for most of the data. We also confirmed that the correlations between the measured $T$ and $R$ fluctuations were negligible.  Any measurements in which $T$ fluctuations gave a comparable or larger contribution than the intrinsic $1/f$ sample noise, such as for $x=0.12$ near its $T_\mathrm{max}$ where $dR/dT$ is high, have been excluded from the analysis (see, e.g., Fig.~2). We note, however, that TLS from the $x=0.12$ sample [Fig.~3(f)] cannot be attributed to $T$ fluctuations, as discussed above.

\section{$1/f$ noise\label{1/f noise}}
\begin{figure*}[!htb]
 \includegraphics[width=0.9\columnwidth]{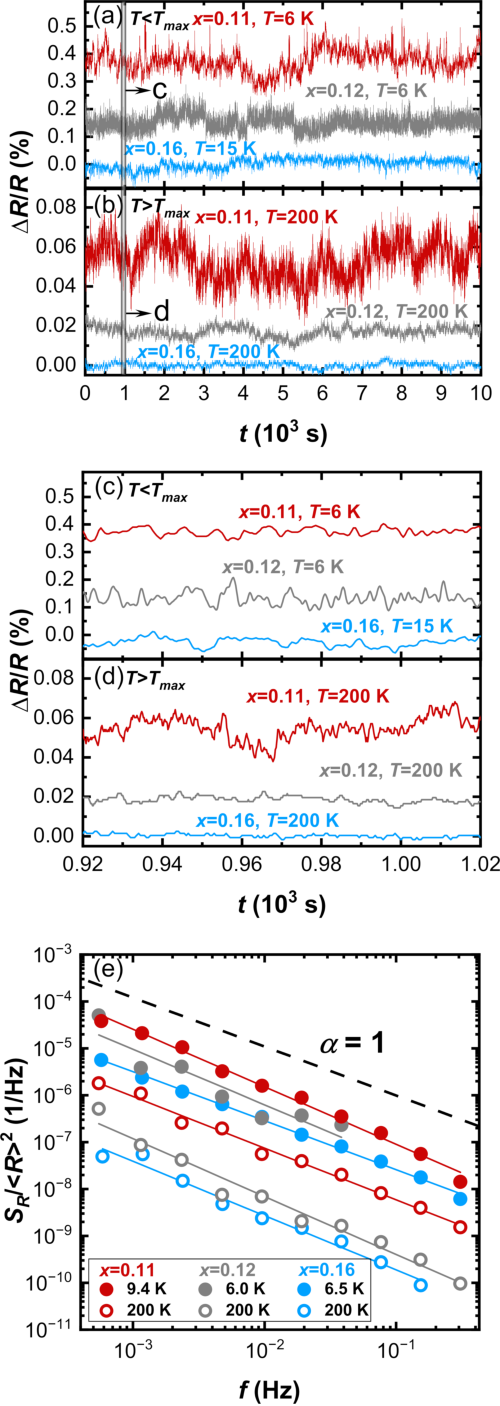}
\caption{Representative $1/f$ noise data from $x=0.11$, $0.12$, and $0.16$ samples (a)--(d) in the time domain and (e) in the frequency domain. In (e), the symbols are octave-averaged power spectra calculated from data in (a) and (b); the solid lines guide the eye.  The dashed line shows the slope for $\alpha=1$.  
\label{fig：1fnoise_supp}}
 \end{figure*}

Some representative $1/f$ noise data are shown in Fig.~S\ref{fig：1fnoise_supp}.
Here we discuss the evolution of the noise magnitude, 
$S_R/\langle R \rangle^2$ at a fixed $f$, with $T$ (Fig.~2). 
At $T/T_\mathrm{max}>1$, the samples are either in or approaching an insulating state (the Mott insulator or the incoherent semiconductor state, which are separated by the quantum Widom line~[S4]). 
In this regime, the simplest possible scenario for the origin of the observed Gaussian $1/f$ noise is that each individual charge carrier acts as an independent fluctuator. As the density of free charge carriers decreases exponentially on lowering $T$, an enhanced $S_R/\langle R \rangle^2$ is expected from Hooge's law~[S5]. At $T/T_\mathrm{max}<1$, where all samples are in the phase-coexistence regime, the noise magnitude for $x=0.11$ demonstrates an exponential growth upon cooling.  A similar but weaker trend is seen for $x=0.12$, while for $x=0.16$, the noise magnitude no longer depends on $T$. Although developing a precise theoretical model to describe the origin of $1/f$ noise is beyond the scope of this work, our results clearly show (Fig.~2) that the noise is strongly enhanced in the regime of phase coexistence ($T/T_\mathrm{max}<1$) compared to that at higher $T$.

\section{Random telegraph noise\label{Additional analyses on RTN}}

\begin{figure*}[!htb]
 \includegraphics[width=2\columnwidth]{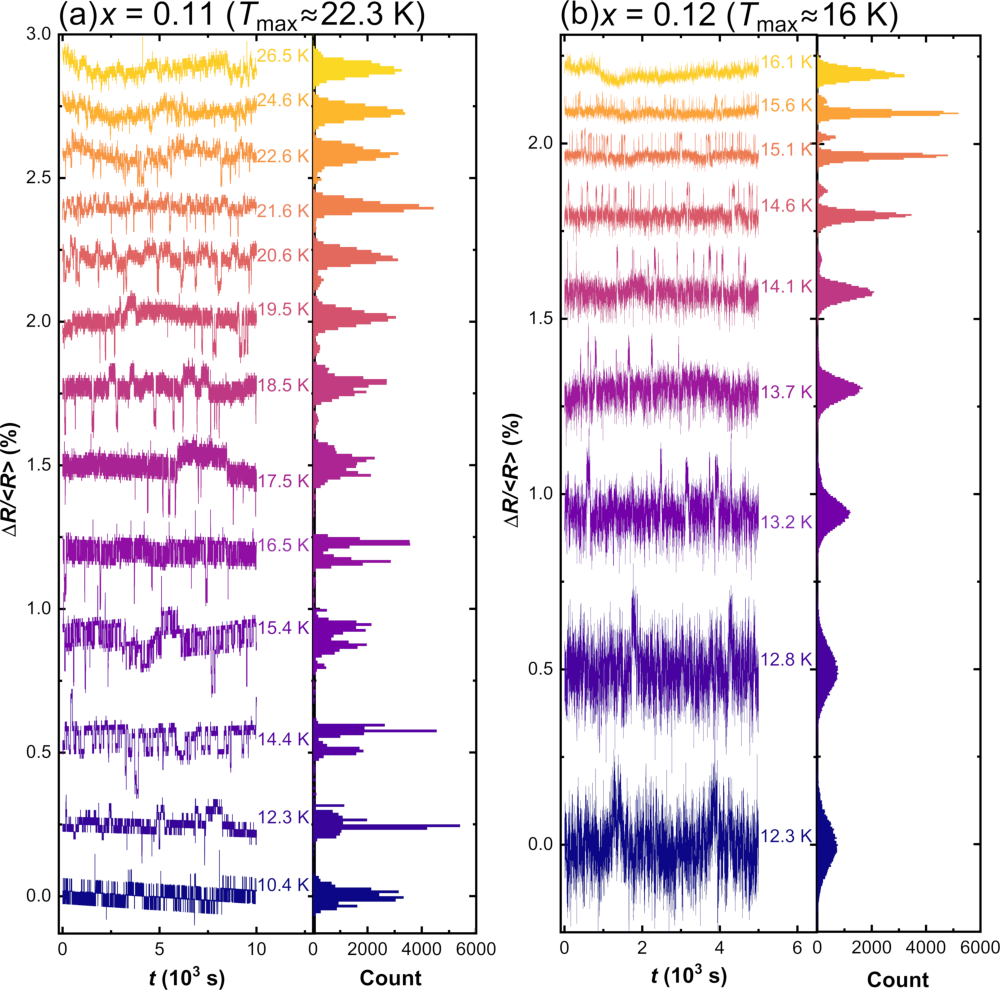}
\caption{The same time-domain traces near $T_\mathrm{max}$ as in (a) $x=0.11$ [Fig.~3(a)] and (b) $x=0.12$ [Fig.~3(f)] and the corresponding histograms that evolve with $T$. The traces are vertically shifted by an arbitrary amount for clarity.
\label{Fig:RTN_histogram}}
\end{figure*}

Figures~S\ref{Fig:RTN_histogram} (a) and (b) show the same time-domain traces $\Delta R/\langle R \rangle$ as in Figs.~3(a) and (f), respectively, with the corresponding histograms of $\Delta R/\langle R \rangle$. As RTN appears, multiple peaks emerge in the histograms.

\begin{figure*}[!htb]
 \includegraphics[width=2\columnwidth]{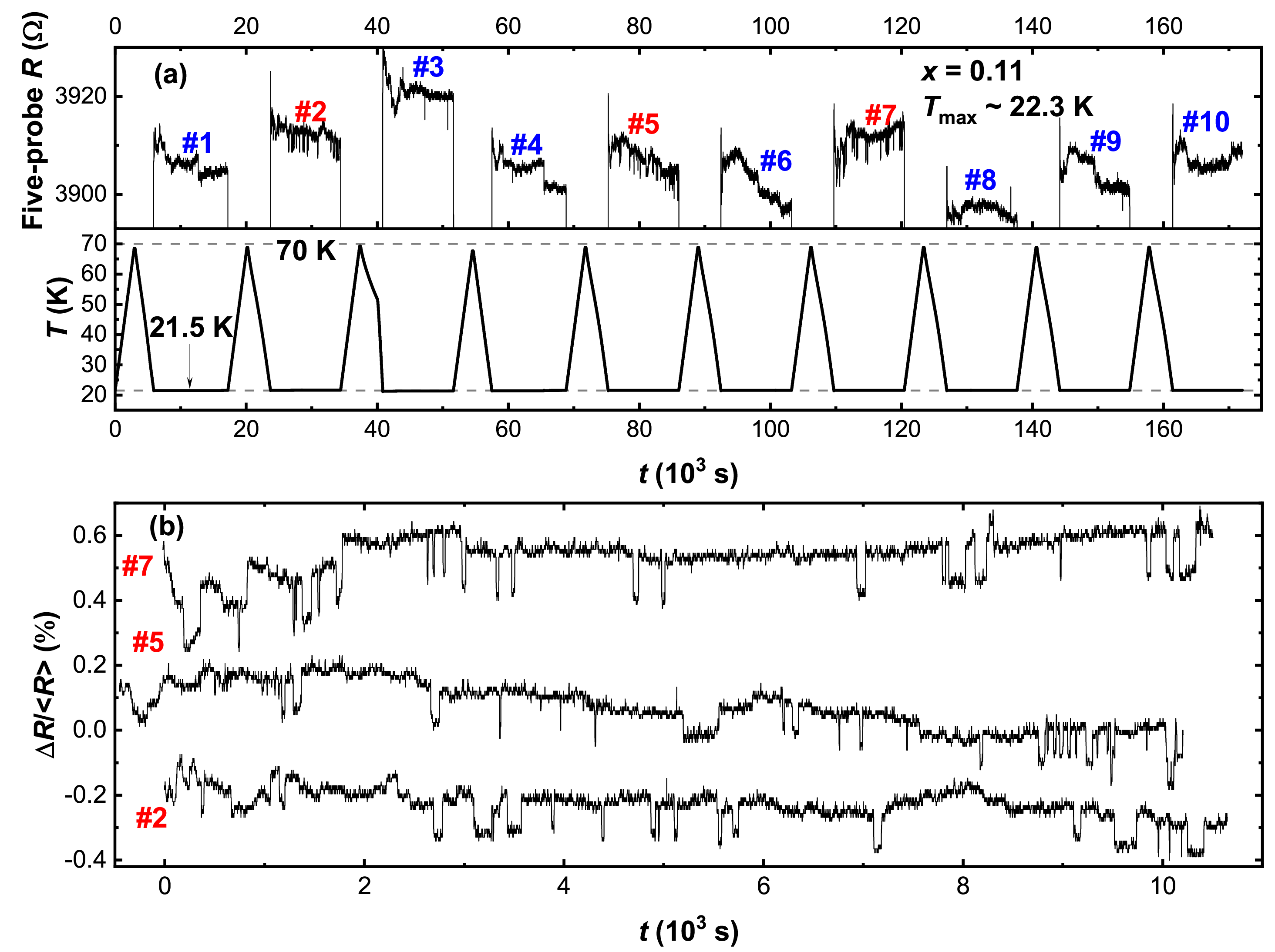}
\caption{(a) Ten consecutive noise measurements from the $x=0.11$ sample at $21.5~\mathrm{K}$, slightly below its $T_\mathrm{max}\sim22.3~\mathrm{K}$. The upper panel shows the resistance measured by a five-probe set-up as a function of time, and the lower panel shows the sample temperature. Well-defined RTN is observed in the second, the fifth, and the seventh measurements. (b) Normalized resistance fluctuations from the three measurements that show RTN; the traces are vertically shifted by an arbitrary amount for clarity.
\label{Fig:Repeat_RTN_011}}
 \end{figure*}

\begin{figure*}[!htb]
 \includegraphics[width=2\columnwidth]{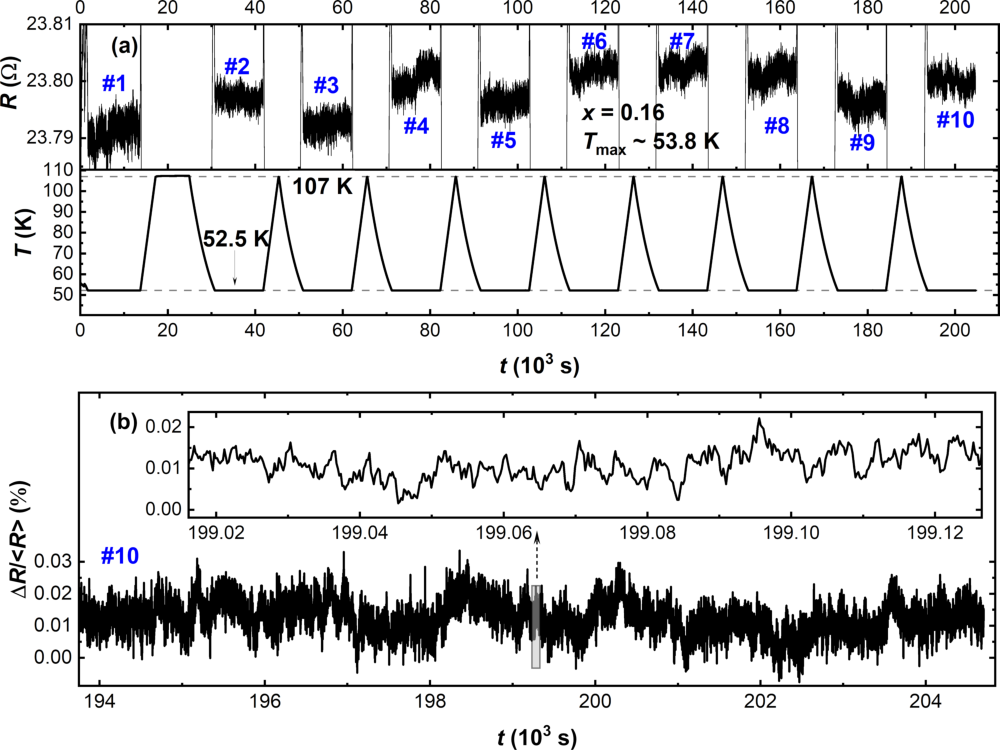}
\caption{(a) Ten consecutive noise measurements from the $x=0.16$ sample at $52.5~\mathrm{K}$, slightly below its $T_\mathrm{max}\sim53.8~\mathrm{K}$. The upper panel shows the resistance as a function of time, and the lower panel shows the sample temperature. No well-resolved RTN is observed. (b) Zoomed-in views for one of the noise measurements.
\label{Fig:Repeat_cycle_016}}
 \end{figure*}

Figure~S\ref{Fig:Repeat_RTN_011} examines the reproducibility of RTN after large thermal cycles. The $x=0.11$ sample is consecutively cycled between $21.5~\mathrm{K}$ and $70~\mathrm{K}$ at $\sim0.85~\mathrm{K/min}$ ten times. Noise was measured every time the sample is cooled down to $21.5~\mathrm{K} \lesssim T_\mathrm{max}\sim22.3~\mathrm{K}$. This temperature ($21.5~\mathrm{K}$) was chosen because, for both $x=0.11$ and $x=0.12$, RTN was observed slightly below their respective $T_\mathrm{max}$ [Figs.~2(a) and (b)]. The thermal cycles and the noise results are presented in Fig.~S\ref{Fig:Repeat_RTN_011}, where three out of ten noise measurements show RTN. For comparison, the $x=0.16$ sample was cycled between $52.5~\mathrm{K}$ and $107~\mathrm{K}$ at $\sim1~\mathrm{K/min}$ for ten consecutive noise measurements at $52.5~\mathrm{K}$, slightly below its $T_\mathrm{max}\sim53.8~\mathrm{K}$, but no well-resolved TLS was observed (Fig.~S\ref{Fig:Repeat_cycle_016}).

\begin{figure*}[!htb]
 \includegraphics[width=1\columnwidth]{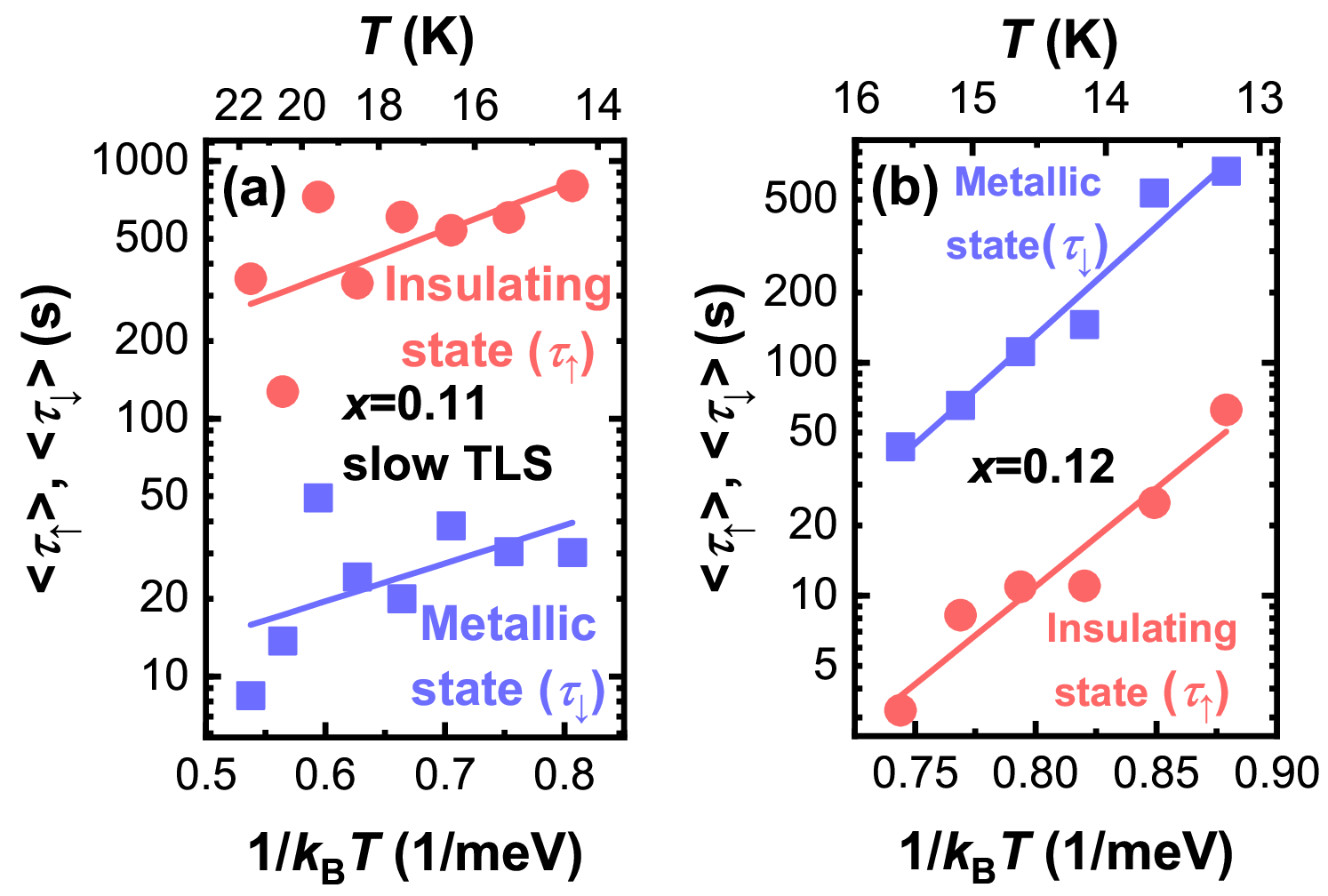}
\caption{Arrhenius plot of the RTN mean lifetime $\langle \tau_i \rangle$ derived from (a) slow TLS of $x=0.11$ data in Fig.~3(a), suggesting $\tau_{0,\uparrow}=31.2\pm11.0~\mathrm{s}$, $\tau_{0,\downarrow}=2.5\pm0.9~\mathrm{s}$, $E_{\uparrow}=4.1\pm2.0~\mathrm{meV}$, and $E_{\downarrow}=3.4\pm2.1~\mathrm{meV}$, and (b) $x=0.12$ data in Fig.~3(f), suggesting $\tau_{0,\uparrow}=2.1\pm0.5~\mu\mathrm{s}$, $\tau_{0,\downarrow}=4.5\pm1.4~\mu\mathrm{s}$, $E_{\uparrow}=19.4\pm2.6~\mathrm{meV}$, and $E_{\downarrow}=21.5\pm2.2~\mathrm{meV}$. The uncertainties are determined from the standard deviation of the fits.
\label{Fig:RTN_012}}
 \end{figure*}

Figure~S\ref{Fig:RTN_012} shows the Arrhenius plot 
of the random telegraph noise lifetime $\langle \tau_i \rangle$
measured from the slow TLS of $x=0.11$ data [Fig.~3(a)] and the $x=0.12$ data [Fig.~3(f)], suggesting thermally activated behavior $\langle \tau_i \rangle=\tau_{0,i}\exp\left( {{E_i}/{k_\mathrm{B}T}}\right)$, similar to Fig.~4(b).

\begin{figure*}[!htb]
 \includegraphics[width=1.2\columnwidth]{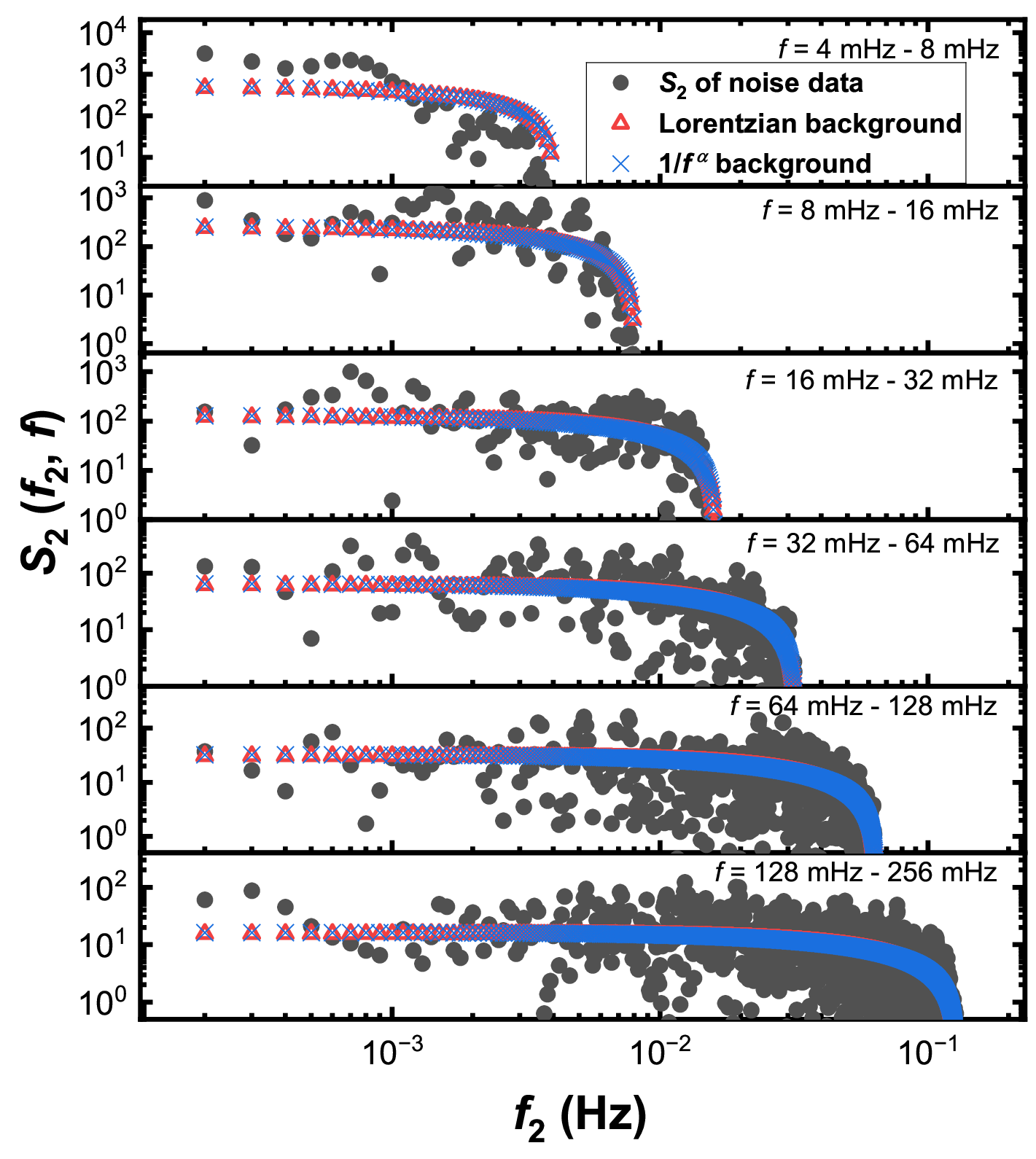}
\caption{The second spectrum of the $x=0.11$, $T=17.5~\mathrm{K}$ noise data from Fig.~3(a). The sub-panels from top to bottom show the second spectrum of different octaves, from $f=4~\mathrm{mHz}$ - $8~\mathrm{mHz}$ to $f=128~\mathrm{mHz}$ - $256~\mathrm{mHz}$. In each sub-panel, the black circles are calculated from the noise data, while the red triangles and the blue crosses are calculated from a stationary Lorentzian spectrum and a $1/f^\alpha$ spectrum, respectively~[S6].
\label{Fig:Second_spectrum}}
\end{figure*}

Figure~S\ref{Fig:Second_spectrum} shows the second spectrum~[S6], a fourth-order noise statistic, of the $x=0.11$, $T=17.5~\mathrm{K}$ noise data presented in Fig.~3(a). Here, the black circles that represent the second spectrum of the noise do not show a distinguishable deviation from the second spectrum of a stationary Lorentzian spectrum or a $1/f^\alpha$ spectrum~[S6]. This result suggests that the RTN arises from a stationary process.

Finally, comparing Figure~S\ref{Fig:RTN_012}, 
and Fig. 3(c), one sees that the energy barrier $E_i$ is always in the order of $10~\mathrm{meV}$, with the free energy difference $|E_\uparrow-E_\downarrow|$ no more than a few $\mathrm{meV}$. 

\vspace{6pt}
\noindent\textbf{{References}}
\newline
\noindent [S1] U. Geiser, H. H. Wang, K. D. Carlson, J. M. Williams, H. A. Charlier Jr., J. E. Heindl, G. A. Yaconi, B. J. Love, M. W. Lathrop, J. E. Schirber, D. Overmyer, J. Ren, and M.-H. Whangbo, Inorg.~Chem.~\textbf{30}, 2586 (1991).
\newline
\noindent [S2] M. Pinteri\'{c}, M. \v{C}ulo, O. Milat, M. Basleti\'{c}, B. KorinHamzi\'{c}, E. Tafra, A. Hamzi\'{c}, T. Ivek, T. Peterseim,
K. Miyagawa, K. Kanoda, J. A. Schlueter, M. Dressel, and
S. Tomi\'{c}, Phys.~Rev.~B.~\textbf{90}, 195139 (2014).
\newline
\noindent [S3] J. H. Scofield, Rev.~Sci.~Instrum.~\textbf{58}, 985 (1987).
\newline
\noindent [S4] A. Pustogow, M. Bories, A. L\"{o}hle, R. R\"{o}sslhuber,
E. Zhukova, B. Gorshunov, S. Tomi\'{c}, J. A. Schlueter,
R. H\"{u}bner, T. Hiramatsu, Y. Yoshida, G. Saito, R. Kato,
T.-H. Lee, V. Dobrosavljevi\'{c}, S. Fratini, and M. Dressel, Nat.~Mater.~\textbf{17}, 773 (2018).
\newline
\noindent [S5]  F. Hooge, Phys.~Lett. A~\textbf{29}, 139 (1969).
\newline
\noindent [S6]  G. T. Seidler and S. A. Solin, Phys.~Rev. B~\textbf{53}, 9753 (1996).

\end{document}